\documentclass[aps,prl,twocolumn,showpacs,lengthcheck,preprintnumbers,a4]{revtex4-1}%
\usepackage{amsfonts}
\usepackage{amsmath}
\usepackage{amssymb}
\usepackage{graphicx}%

\begin{document}
\preprint{\normalsize Scientific Reports \textbf{8}, 10940 (2018)}
%\preprint{Scientific Reports \textbf{8}, 10940 (2018)}
\title{\Large Stable symmetry-protected 3D embedded solitons in Bose--Einstein condensates}

\author{V. Delgado$^1$} 
\email{vdelgado@ull.es} 
\author{A. Mu\~{n}oz Mateo$^2$}
%\email{ammateo@ull.es}

\affiliation{$^1$ Departamento de F\'{\i}sica, Facultad de Ciencias,
Universidad de La Laguna, La Laguna, Tenerife, Spain}
\affiliation{$^2$ Dodd-Walls Centre for Photonic and Quantum
Technologies and Centre for Theoretical Chemistry and Physics, New Zealand
Institute for Advanced Study, Massey University, %Private Bag 102904 NSMC, 
Auckland 0745, New Zealand}

\begin{abstract}
Embedded solitons are rare self-localized nonlinear structures that,
counterintuitively, survive inside a continuous background of resonant states.
While this topic has been widely studied in nonlinear optics, it has received almost 
no attention in the field of Bose--Einstein condensation. 
In this work, we consider experimentally realizable Bose--Einstein condensates
loaded in one-dimensional optical lattices and demonstrate that they support 
\emph{continuous} families of \emph{stable} three-dimensional (3D) embedded solitons.
These solitons can exist inside the resonant continuous Bloch bands because they are 
protected by symmetry.
The analysis of the Bogoliubov excitation spectrum as well as the long-term evolution 
after random perturbations proves the robustness of these nonlinear structures against 
any weak perturbation. This may open up a way for the experimental realization of 
stable 3D matter-wave embedded solitons as well as for monitoring the 
gap-soliton to embedded-soliton transition.
\end{abstract}
\maketitle

%\thispagestyle{empty}

%\noindent

%\section*{Introduction}

Solitons have attracted much interest since their discovery in 1834. %by Scott-Russell. 
They are a peculiar manifestation of nonlinear wave systems resulting from a 
detailed balance between dispersion and nonlinearity \cite{Kivshar2003}. 
Solitons are ubiquitous in Physics, where they appear as particle-like
self-localized coherent solutions of certain differential equations 
such as the well-known nonlinear Schr\"odinger equation.
This equation has great relevance in the field of nonlinear optics, where it 
governs the propagation of the envelope of an electromagnetic pulse \cite{Kivshar2003}
and plays also an essential role in the field of Bose--Einstein condensation \cite{CGFK}, 
where it adopts the form of the Gross--Pitaevskii equation (GPE) and describes the 
dynamics of Bose--Einstein condensates (BECs) of dilute atomic gases 
\cite{Pitaev,Pethick}. 
Not surprisingly, in both the above fields the study of solitons have received 
great attention in recent years. In nonlinear optics, solitons
have found broad practical application in telecommunications via optical 
fibers. 
In Bose--Einstein condensation, dark solitons have been experimentally realized in 
atomic condensates with repulsive interparticle interactions
\cite{Burger1999,Denschlag2000}.
These solitons represent the matter-wave 
counterpart of optical dark solitons in self-defocusing nonlinear media.
Matter-wave bright solitons have been obtained in attractive condensates 
(the equivalent of self-focusing nonlinear media)
\cite{Strecker2002,Khaykovich2002}.
The matter-wave analogue of optical gap solitons has also been experimentally 
observed in repulsive condensates of $^{87}$Rb atomic gases
\cite{Eiermann2004}.
Matter-wave gap solitons are self-trapped nonlinear structures that can be found
in BECs immersed in the periodic potential of an optical lattice
\cite{Louis2003,Morsch2006}.
They are localized nonlinear stationary states of the GPE whose chemical 
potentials lie in the forbidden band gaps of the linear spectrum.
In a repulsive condensate, gap soliton solutions bifurcate from the upper edge
of the linear Bloch bands, forming distinct one-parameter families characterized
by their chemical potential $\mu$ as a function of the number of atoms $N$
\cite{VDB2011,VDB2014}.
These families describe continuous trajectories in the $\mu$--$N$ parameter plane
and, in general, vanish as they approach the vicinity of an upper 
band, where the gap solitons become resonant with the extended Bloch waves
residing therein. In these regions, in general, only delocalized solutions with 
nonvanishing oscillating tails can exist.
However, under certain circumstances localized solitons can be found
inside the continuous background of resonant states \cite{Champneys2001}. 
These solitons, that reside in the continuous spectrum, were called
embedded solitons and studied in Ref. \cite{Yang1999} in the 
framework of nonlinear optics.
Both isolated (at specific parameter values) and continuous families of embedded 
solitons have been reported in the literature and, usually, they were found to 
be semistable \cite{ Champneys2001,Yang1999,Yang2003,Yang2003a}. 
Most studies have been restricted to the one-dimensional case and finding
higher dimensional embedded solitons still remains a challenge. In this regard,
continuous families of fully stable 2D embedded solitons were obtained in 
Ref. \cite{Yang2010} in a self-defocusing optical medium in a quasi-1D waveguide 
array characterized by a refractive index variation
\begin{equation}
n(x,y)=6\,e^{-y^2/4} \cos^2 x.
\label{nxy}%
\end{equation}
Symmetry plays an important role in the existence of these solitons, which can 
only exist inside Bloch bands with opposite parity from the solitons themselves
\cite{Yang2010}.

Embedded solitons have received almost no attention in the field of Bose--Einstein
condensation. This may be, in part, because most theoretical studies on gap solitons 
have focused on quasi-1D BECs \cite{Louis2003,VDB2011,VDB2014,Salasnich}. 
Nonetheless, 3D gap waves \cite{Alexander2006} (a type of gap solitons that can be 
viewed as truncated nonlinear Bloch waves) and 3D gap solitons \cite{VDB1013} have 
been obtained, respectively, in BECs loaded in 3D and in 1D optical lattices.  
On the other hand,
the authors of Ref. \cite{Huang2018} have studied embedded solitons using a 
generalized 2D GPE that incorporates an extra momentum operator of the form 
$-i (\gamma_x \partial_x+ \gamma_y \partial_y)$, where $\gamma_x$ and $\gamma_y$ 
are adjustable parameters. They found that in the presence of a periodic potential 
$V(x,y)$ of the same functional form as the refractive index $n(x,y)$ of 
Eq. (\ref{nxy}), this theoretical model admits stable 2D embedded solitons which, as 
occurs with those previously found in Ref. \cite{Yang2010}, can only
exist inside Bloch bands with opposite symmetry to that of the solitons.

Despite all the efforts, no stable 3D embedded solitons have been obtained so far 
neither in nonlinear optics nor in BECs. In this work, we obtain for the first time stable 
3D embedded solitons. Specifically, we consider experimentally realizable 
single-component BECs subject to a sufficiently weak transverse confinement and 
loaded in 1D optical lattices and demonstrate that they support \emph{continuous} families 
of \emph{stable} 3D gap-solitons that can survive inside the continuous background 
of resonant Bloch waves.
Since, under these circumstances, the GPE commonly does not admit localized
stationary solutions, the existence of such embedded solitons requires the fulfillment 
of certain special conditions.
The question arises as to why these embedded solitons can exist. 
To address this question, we perform a realistic 3D numerical treatment that
enables us to follow the gap-soliton to embedded-soliton transition and to understand 
the physics behind their existence.  %of the above embedded solitons.
Our numerical results indicate that these solitons can exist inside the resonant 
Bloch bands because they are protected by rotational symmetry. 
By calculating the Bogoliubov excitation spectrum and the long-term 
nonlinear evolution after random perturbations, we have checked that these 
solitons are fully stable, which may open up a way for experimentally generating robust 
3D embedded solitons in BECs and for investigating the corresponding gap-soliton to 
embedded-soliton transition.

%\section*{Model}
%\vspace*{0.8cm}
\vspace*{22pt}
{\noindent \Large \textbf{Model} \vspace*{1pt}}

\noindent
In this work, we consider a $^{87}$Rb condensate subject to a transverse confinement 
of frequency $\omega_{\bot}/2\pi=320$ Hz and loaded in a 1D optical lattice of period 
$d=\pi/2$ $\mu$m and depth $s=15$ in units of the recoil energy
$E_{R}\equiv\hbar^{2}\pi^{2}/2m_0 d^2=0.75\hbar\omega_{\bot}$ (where $m_0$ is the atomic 
mass). Its dynamics is governed by the 3D GPE
\begin{equation}
i\hbar\partial_t\psi=\left(  -\hbar^{2}\boldsymbol{\nabla}^{2}/{2m_0}
+V(\mathbf{r})+gN\left\vert \psi\right\vert ^{2}\right)  \psi,
\label{3DGPE}%
\end{equation}
which accurately describes the physics of zero-temperature BECs in the mean 
field regime under realistic experimental conditions
\cite{VDB2006,Mottonen}. In this equation,
$V(\mathbf{r})=\frac{1}{2}m_0\omega_{\bot}^2\mathbf{r}_{\bot}^2+s E_{R}\sin^{2}(\pi z/d)$ 
is the external potential, $N$ is the number of atoms in the condensate, 
and $g=4\pi\hbar^{2}a/m_0$ is the interatomic interaction 
strength, with $a=5.29$ nm being the s-wave scattering length.
To determine the spectrum of the underlying linear problem, we note that the 
Hamiltonian is separable. The transverse eigenvalue equation can be solved analytically. 
Its solutions are the eigenmodes of the radial harmonic oscillator
\begin{equation}
\varphi_{n_{r}}^{(m)}(\rho,\theta)=
\sqrt{\frac{n_{r}!}{\pi a_{\bot}^{2}(n_{r}+|m|)!}}
e^{im\theta}\rho
^{|m|}e^{-\rho^{2}/2}L_{n_{r}}^{(|m|)}(\rho^{2}), \label{Eq-2}%
\end{equation}
with corresponding eigenvalues
\begin{equation}
E_{m,n_r}=(2n_{r}+|m|+1)\hbar\omega_{\bot}. \label{Eq-3}%
\end{equation}
$L_{n_{r}}^{(|m|)}(\rho^{2})$ are generalized Laguerre 
polynomials, with $\rho\equiv r_{\bot}/a_{\bot}$ where 
$a_{\bot}=\sqrt{\hbar/m_0\omega_{\bot}}$ is the harmonic-oscillator length
and $(r_{\bot},\theta)$ are the polar coordinates; 
$m=0,\pm1,\pm2,\ldots$ is the axial angular momentum quantum number, and 
$n_{r}=0,1,2,\ldots$ is the radial quantum number.
Thus, the problem reduces to solving the 1D axial eigenvalue equation. This is a stationary 
Schr\"odinger equation in a periodic potential whose solutions are Bloch waves
%\begin{equation}
$\phi_{n,q}(z)=e^{iqz}u_n(z),$
%\end{equation}
characterized by the band index $n$ and the quasimomentum $q$.
%in the first Brillouin zone $(-\pi/d,\pi/d\,]$. 
The resulting 3D spectrum reproduces the band-gap structure of the 1D axial 
problem for every $(m,n_r)$ transversal state and 
consists of a series of 3D Bloch bands determined by the 
quantum numbers $(n,m,n_r)$. 

%%%%%%%%%%%%%%%%%%%%%%%%%%%%%%%%%%%%%%%%%%%%%%%%%%%%%%%%%%%%%%%%%%%%%%%%%%%%%%%

\begin{figure}[tpb]
\begin{center}
\includegraphics[
width=9.0cm
]{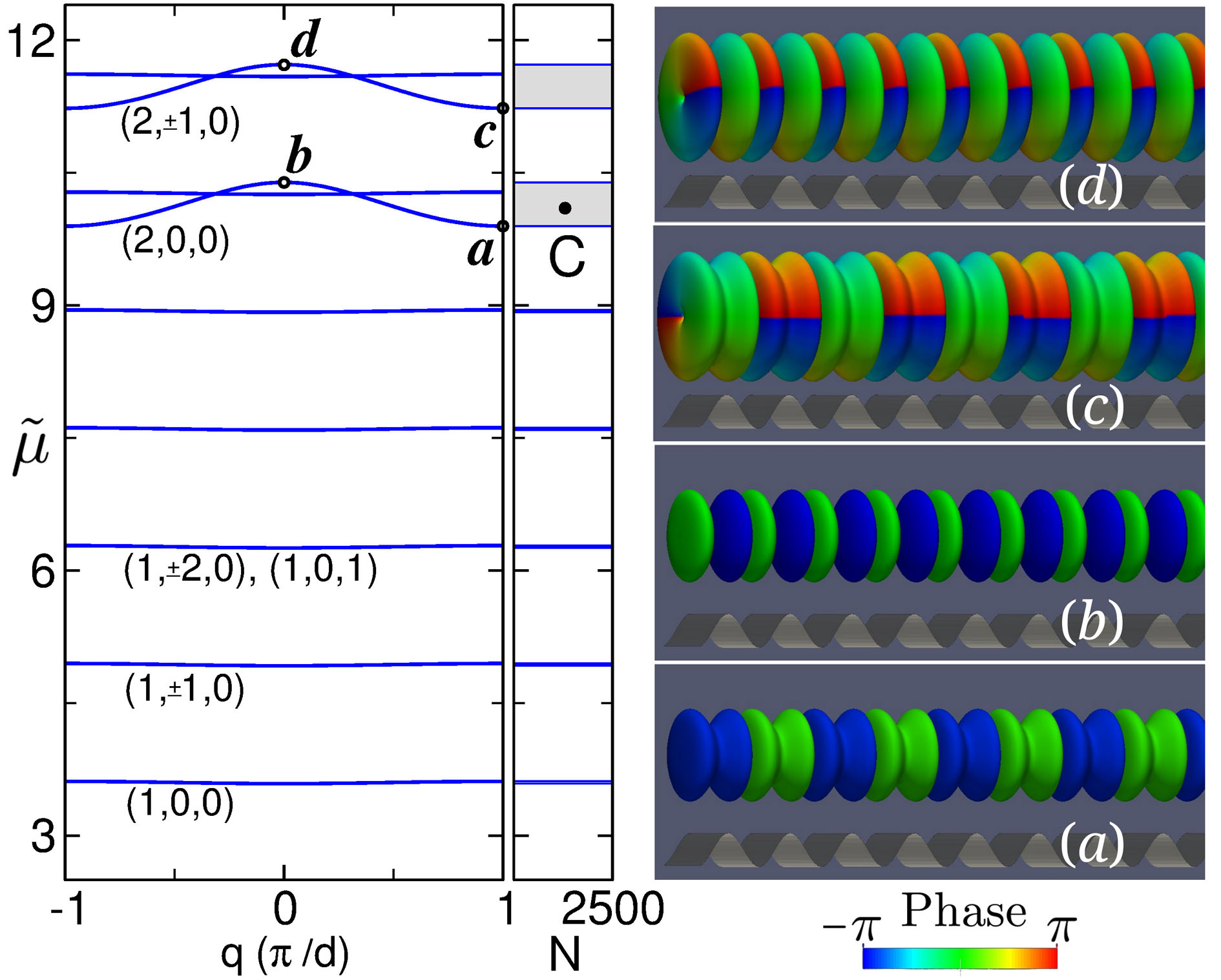}
\end{center}
\caption{
Left: Ideal-gas ($gN=0$) linear spectrum. Point C corresponds to an embedded soliton 
with $N=1130$.
Right: Phase-colored density isosurfaces of the Bloch waves corresponding to the points
labeled with the same letters on the left.}%
\label{Fig1}%
\end{figure}

%%%%%%%%%%%%%%%%%%%%%%%%%%%%%%%%%%%%%%%%%%%%%%%%%%%%%%%%%%%%%%%%%%%%%%%%%%%%%%%

This spectrum is depicted in Fig. \ref{Fig1}, which shows the scaled chemical potential 
$\widetilde{\mu}\equiv(\mu-\hbar\omega_{\bot})/E_{R}$ as a function of $q$. 
For a given band index $n$, there exists an infinite series of replicas of the
lowest-energy Bloch band $(n,0,0)$ corresponding to the different $(m,n_r)$
excited transversal states \cite{VDB2011,VDB1013}. 
Panels ($a$)--($d$) in Fig. \ref{Fig1},
show the Bloch waves corresponding to the points
labeled with the same letters on the left,
%the band-gap spectrum of Fig. \ref{Fig1}, 
represented as phase-colored 
% density 
isosurfaces of the atom density taken at 5\% of its maximum. The optical lattice is 
also shown for reference at the bottom.
%%%%%%%%%%%%%%%%%%%%%%%%%%%%%%%%%%%%%%%%%%%%%%%%%%%%%%%%%%%%%%%%%%%%%%%%%%%%%%%
%
\begin{figure*}[tpb]
\begin{center}
\includegraphics[
width=17.5cm
]{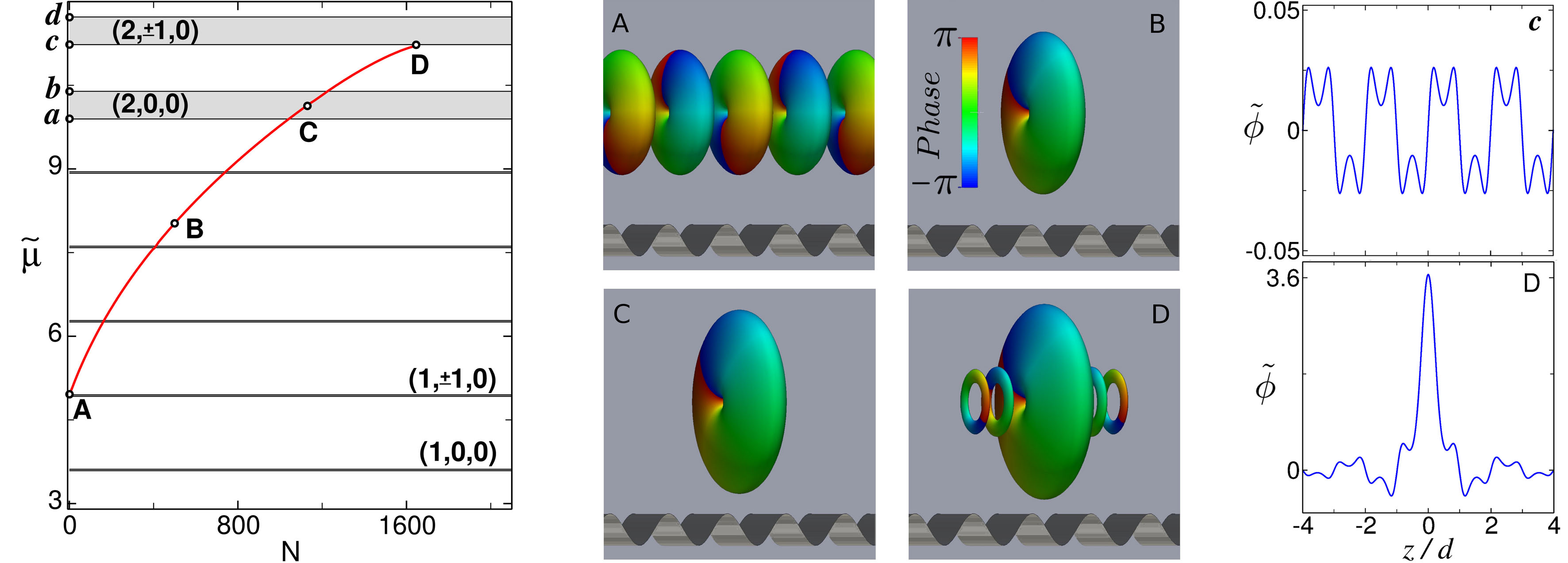}
\end{center}
\caption{
Left: Trajectory in $\mu$--$N$ plane of the $(1,1,0)$ soliton family.
Middle: Phase-colored density isosurfaces corresponding to the points
labeled with the same letters on the left panel.
Right: Axial profiles of the 3D wave functions corresponding to the points 
$c$ and $D$ on the leftmost panel. Note that panel $c$ refers to the wave function 
shown in Fig. \ref{Fig1}($c$) while panel $D$ refers to the wave function shown in
the middle panel $D$ of the present figure.
}%
\label{Fig2}%
\end{figure*}
%
%%%%%%%%%%%%%%%%%%%%%%%%%%%%%%%%%%%%%%%%%%%%%%%%%%%%%%%%%%%%%%%%%%%%%%%%%%%%%%%
Panels ($a$) and ($b$) correspond to the minimum and maximum chemical potential 
of the $(2,0,0)$ band, while panels ($c$) and ($d$) correspond to the 
%minimum and maximum chemical potential
equivalent points of the $(2,1,0)$ band. Both of these bands
originate from the first excited Bloch band of the 1D axial problem.
Bloch waves in this excited 1D band consist of 
two out-of-phase axial peaks (one axial node) at every lattice site.
The main difference between the eigenmodes displayed in the figure is 
that while Bloch modes in the $(2,0,0)$ band are in their (topologically trivial) 
transverse ground state, those in the $(2,1,0)$ band exhibit a central vortex 
with topological charge $m=1$.

Because of the axial symmetry of  the Hamiltonian $H$, eigenstates with different 
values of $|m|$ (which belong to different irreducible invariant subspaces of the 
rotation group) cannot be dynamically connected by $H$.
As we will see, this fact remains true in the nonlinear regime and ultimately 
allows for the existence of symmetry-protected embedded solitons inside the resonant 
Bloch bands. Point $C$ in Fig. \ref{Fig1} marks the location of one of these embedded 
solitons with $N=1130$.

While in the linear regime all the stationary states of the GPE in an optical lattice are 
extended (delocalized) Bloch waves, in the nonlinear regime there also exist stationary 
solutions in the form of families of self-localized gap solitons residing within the band gaps.
To obtain these gap soliton families, we numerically seek stationary 
solutions of the form
%\begin{equation}
$\psi(\mathbf{r},t)=\psi_0(\mathbf{r})\exp(-i\mu t/\hbar),$ %\label{Eq-4}%
%\end{equation}
where the chemical potential $\mu$ is used as the continuation
parameter for a Newton continuation method implemented in terms of a 
Laguerre--Fourier spectral basis consisting of the eigenfunctions 
$\varphi_{n_{r}}^{(m)}(\rho,\theta)$ of the radial harmonic oscillator, 
Eq. (\ref{Eq-2}), along with the plane-wave eigenfunctions of the axial 
linear momentum $P_z=-i \hbar \partial_z$. 
As the continuation parameter $\mu$ increases (or equivalently, as $N$ 
increases), gap soliton families describe distinctive trajectories
$\mu(N)$ in the $\mu$--$N$ plane.
Figure \ref{Fig2} displays the trajectory characterizing the family of fundamental 
gap solitons of type $(1,1,0)$, where, following the notation introduced in 
Refs. \cite{VDB2011,VDB1013}, we are denoting gap soliton families by the 
quantum numbers of the linear Bloch band from which they bifurcate. 
The middle panels show the 3D stationary wave functions of points $A$--$D$ in
terms of phase-colored isosurfaces of the atom density, while the
right panels depict the axial profiles (obtained by integrating out the radial coordinate 
for a fixed $\theta$) of the 3D wave functions of points $c$ and $D$ (see Figs. \ref{Fig1}(c)
and \ref{Fig2}(D), respectively).
In the linear regime (point $A$), the stationary solution in this family is 
an extended Bloch wave featuring an axisymmetric vortex with topological 
charge $m=1$ (panel $A$). Note the $\pi$-phase shift along the $z$ axis for
every given $\theta$ in addition to the 2$\pi$-phase shift around the vortex
singularity. 
As already said, in the nonlinear regime ($gN>0$), localized stationary solutions 
appear inside the band gaps of the linear spectrum. These solutions, which bifurcate 
from the upper edge of the Bloch band giving rise to the continuous $\mu(N)$ family 
shown in the figure, become more localized as one moves away from the Bloch band.
A representative example of a gap soliton in this family (point $B$) is shown in 
Panel $B$. As can be seen, it is a gap soliton with topological charge $m=1$,
fully localized inside a single potential well. 
All solitons in this family are eigenstates of the axial angular momentum $L_z$
with eigenvalue $m=1$.

%%%%%%%%%%%%%%%%%%%%%%%%%%%%%%%%%%%%%%%%%%%%%%%%%%%%%%%%%%%%%%%%%%%%%%%%%%%%%%%

\begin{figure}[t]
\begin{center}
\includegraphics[
width=8.5cm
]{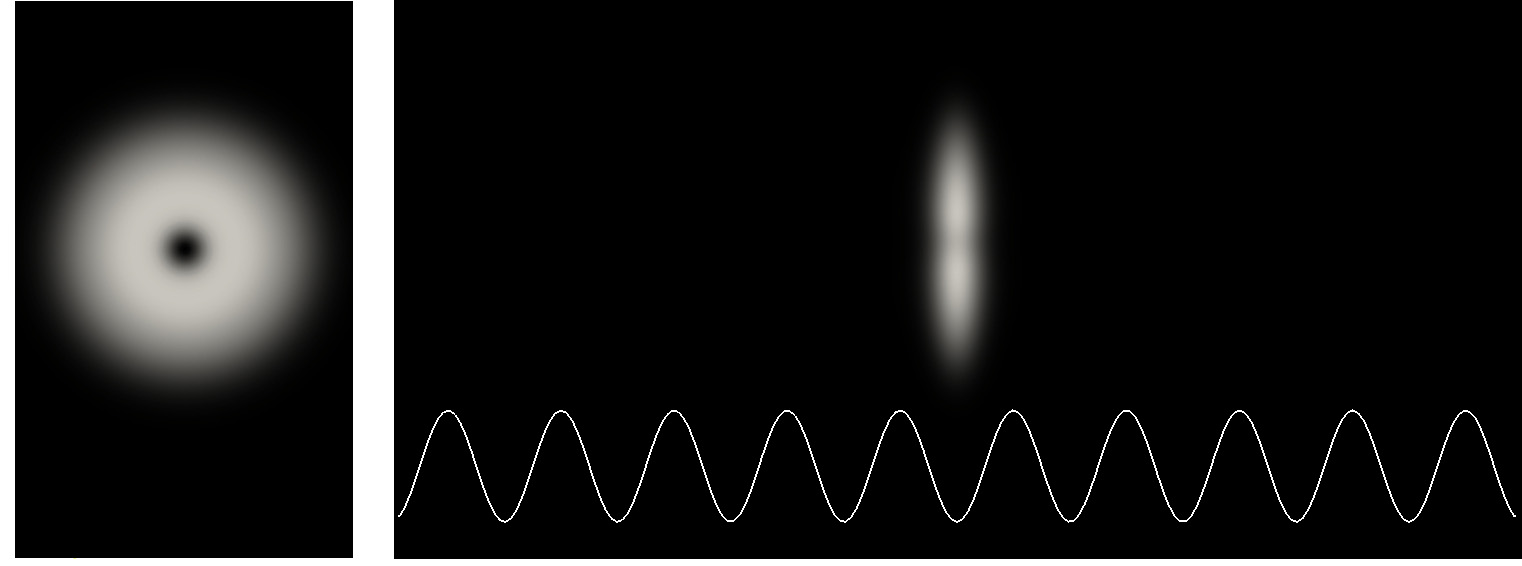}
\end{center}
\caption{
Column densities of the embedded soliton $C$ of Fig. \ref{Fig2} integrated along the axial
(left panel) and transverse (right panel) directions.}%
\label{Fig3}%
\end{figure}

%%%%%%%%%%%%%%%%%%%%%%%%%%%%%%%%%%%%%%%%%%%%%%%%%%%%%%%%%%%%%%%%%%%%%%%%%%%%%%%

Most commonly, characteristic trajectories $\mu(N)$ of gap soliton families vanish 
as they enter Bloch bands. This occurs because inside the allowed energy bands, 
gap solitons resonantly couple with the extended linear Bloch waves,
which eventually prevents the existence of localized stationary states.
However, as can be seen in Fig. \ref{Fig2}, gap solitons in this family survive inside 
the $(2,0,0)$ Bloch band. %(see also Fig. \ref{Fig1}). 
These surviving solitons, with chemical
potential $\widetilde{\mu}$ between points $a$ and $b$, represent a continous $\mu(N)$ 
family of symmetry-protected 3D embedded solitons. 
Point $C$ in Figs. \ref{Fig1} and \ref{Fig2} constitutes a representative example
of one of these solitons. As can be seen in panel $C$, this soliton is a genuine 
self-localized 3D embedded soliton with topological charge $m=1$.
Its existence is a direct consequence of the rotational symmetry of the system, which
forbids the solitons in this family $|\psi_m\rangle$ to couple 
with resonant linear 
Bloch waves $| \phi_{m'}\rangle$ with different values of $m$.
Indeed, taking into account that $|\psi_m\rangle$ are eigenstates of the nonlinear 
hamiltonian,
$H|\psi_m\rangle=\mu |\psi_m\rangle$,
and that both $|\psi_m\rangle$ and $| \phi_{m'}\rangle$ are eigenstates of $L_z$,
one finds 
$\langle \phi_{m'} |H|\psi_m\rangle=\mu \langle \phi_{m'} |\psi_m\rangle \propto \delta_{m'm}$.
Thus, $H$ cannot dynamically couple soliton families with Bloch waves having
different rotational properties. They are protected by symmetry.
As a result, fundamental solitons of the $(1,1,0)$ family can safely reside within
the $(2,0,0)$ Bloch band.
On the contrary, as is apparent from the figure, no localized $(1,1,0)$ soliton solutions
can be found inside the $(2,1,0)$ spectral band. %(which extends between points $c$ and $d$).
In fact, as gap solitons in this family approach the lower edge of the Bloch 
band (point $D$), they develop an oscillating tail reminiscent of the Bloch waves
residing in the spectral band. 
This can be seen from the middle panel $D$ in Fig. \ref{Fig2}, that shows the density isosurface of
the gap soliton corresponding to point $D$. As is apparent, in this case the gap soliton 
appears flanked on each side
by a density pattern that exhibits the characteristic features of the 
Bloch waves displayed in Fig. \ref{Fig1}($c$) (two out-of-phase axial peaks (one axial node) at 
each lattice well and a unit-charge central vortex).
Since in isosurface plots the atom density is abruptly cut off, it can be more illustrative
to consider the two rightmost panels $c$ and $D$ of Fig. \ref{Fig2}, which depict, respectively, 
the axial profiles of the 3D wave functions shown in Fig. \ref{Fig1}($c$) and in the middle panel 
$D$ of Fig. \ref{Fig2}. A simple comparison reflects that in this case, in the vicinity of the Bloch 
band, gap solitons become contaminated by the extended Bloch waves residing therein.

In Fig. \ref{Fig3} we show the axial and transverse column densities of the embedded soliton 
$C$ of Fig. \ref{Fig2}, which are the quantities directly measured in experiments.

We have also investigated the family of fundamental $(1,0,0)$ gap solitons.
Besides bifurcating from the lowest Bloch band and being, consequently, in their
topologically trivial ground state, the numerical results obtained for 
the solitons in this family are qualitatively similar to those of Fig. \ref{Fig2} and are not explicitly 
shown.
As expected, solitons in this family survive protected by symmetry
inside the $(2,\pm 1,0)$ 
Bloch band, while they do not exist in the vicinity of the $(2,0,0)$ band.
As we will see, however, these $(1,0,0)$ embedded solitons have different 
stability properties than the topologically protected $(1,1,0)$ solitons considered above.

It is worth remarking the crucial role that the dimensionality of the problem plays in the 
existence and properties of the embedded solitons considered in this work.
While certain properties of gap solitons, such as the chemical potential $\mu(N)$, can be
obtained in an approximate way in terms of an effectively 1D model, the very existence of
the above embedded solitons is a direct consequence of a specific interplay between
axial and transverse degrees of freedom, which clearly cannot be accounted for by any
1D model. This is a genuine 3D system in which transverse and axial degrees of
freedom play an equally relevant role that must be explicitly incorporated in the description 
of the problem.

%\section*{Stability}
%\vspace*{0.8cm}
\vspace*{22pt}
{\noindent \Large \textbf{Stability} \vspace*{1pt}}

\begin{figure}[ptb]
\begin{center}
\includegraphics[
width=8.cm
]{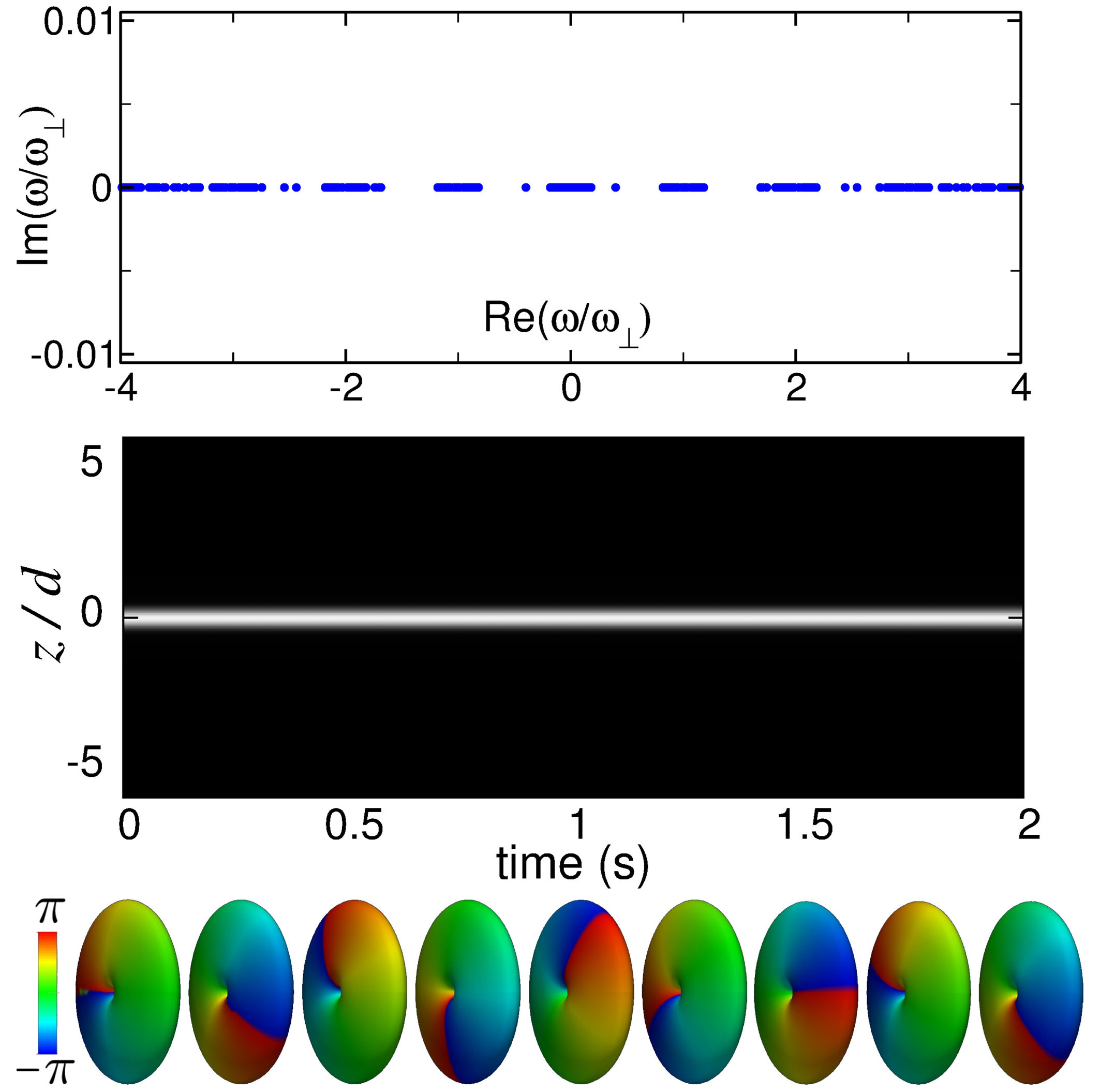}
\end{center}
\caption{
Results of the stability analysis of the embedded soliton $C$ of Fig. \ref{Fig2}.}%
\label{Fig4}%
\end{figure}

\noindent
Dynamical stability is a necessary prerequisite for the above
embedded solitons to have any physical relevance. To investigate this issue,
we begin by performing a linear stability analysis based on the
Bogoliubov spectrum of elementary excitations. To this end, we perturb
the stationary wavefunctions $\psi_{0}(\mathbf{r})$ of the embedded solitons
by introducing a small fluctuation of frequency $\omega$
\begin{equation}
\psi(\mathbf{r},t)=\left[  \psi_{0}(\mathbf{r})+u(\mathbf{r})e^{-i\omega
t}+v^{\ast}(\mathbf{r})e^{i\omega t}\right]  e^{-i\mu t/\hbar}.\label{BdG1}%
\end{equation}
After substituting in the GPE and retaining up to linear terms in the 
amplitudes $u$ and $v$, one obtains the 
corresponding Bogoliubov--de Gennes (BdG) equations
\begin{equation}
\left(
\begin{array}
[c]{cc}%
\mathcal{L} & gN\psi_{0}^{2}(\mathbf{r})\\
-gN\psi_{0}^{\ast 2}(\mathbf{r)} & \mathcal{-L}%
\end{array}
\right)  \left(
\begin{array}
[c]{c}%
u(\mathbf{r})\\
v(\mathbf{r})
\end{array}
\right)  =\hbar\omega\left(
\begin{array}
[c]{c}%
u(\mathbf{r})\\
v(\mathbf{r})
\end{array}
\right),\label{BdG2}%
\end{equation}
where 
%\[
%\begin{equation}
$\mathcal{L}=-(\hbar^{2}/2m_0)\boldsymbol{\nabla}^{2}+V(\mathbf{r}%
)-\mu+2gN\left\vert \psi_{0}\right\vert ^{2}$. 
%\end{equation}
%\]
The solution of the above linear eigenvalue problem provides the desired information.
We have numerically solved Eq. (\ref{BdG2}) by expanding its eigenfunctions in terms 
of the above Laguerre--Fourier spectral basis.
Figure \ref{Fig4} collects the results of the stability analysis of the embedded soliton C
of Fig. \ref{Fig2}. As can be seen in the upper panel, that shows the spectrum of elementary 
excitations, all the BdG eigenfrequencies are real, which demonstrates the linear 
stability of this soliton. 
To investigate the stability in the nonlinear regime, we have produced a random 
perturbation in the stationary wavefunction of the soliton by adding
a small-amplitude ($\sim1\%$) Gaussian white noise and have analyzed its
subsequent nonlinear evolution. 
To this end, we have numerically integrated the 3D GPE by using a pseudospectral
method based on the above Laguerre--Fourier spectral basis along with a
third-order Adams--Bashforth scheme for the time evolution.
The middle panel in Fig. \ref{Fig4} shows the long-time behavior (up to $t=2$ s) of the 
axial profile of the perturbed soliton
by means of a density map where bright pixels indicate high densities.
Below this panel, and using the same time axis, the corresponding 3D wave functions
are displayed as phase-colored density isosurfaces. As is apparent, apart from
the expected stationary-state global phase evolution, the embedded soliton 
remains unaltered, which demonstrates its stability.

%%%%%%%%%%%%%%%%%%%%%%%%%%%%%%%%%%%%%%%%%%%%%%%%%%%%%%%%%%%%%%%%%%%%%%%%%%%%%%%

%%%%%%%%%%%%%%%%%%%%%%%%%%%%%%%%%%%%%%%%%%%%%%%%%%%%%%%%%%%%%%%%%%%%%%%%%%%%%%%

It is interesting to note that stable $(1,1,0)$ embedded solitons can also be found
in the narrow $(1,m,n_r)$ Bloch bands with $|m| \neq 1$, while they disappear
inside the $|m|=1$ resonant bands.
In general, gap and embedded solitons in this $(1,1,0)$ family are stable along
the entire $\mu(N)$ trajectory except in the close proximity to the $|m|=1$ resonant 
bands where they become unstable and eventually disappear as they enter the 
band. 
The soliton $D$ of Fig. \ref{Fig2} is a representative example of one of these 
unstable gap solitons.

On the other hand, while the symmetry-protected $(1,0,0)$ solitons 
embedded in the $(2,\pm 1,0)$ Bloch band are linearly stable against weak 
perturbations that respect the rotational symmetry of the Hamiltonian, they are not, 
however, fully stable. 
% The analysis of the BdG spectrum reveals the existence of 
% nonsymmetric elementary excitations with complex eigenfrequencies.
Indeed, the analysis of the BdG spectrum reveals that all the rotationally symmetric 
elementary excitations have real eigenfrequencies, while there exist nonsymmetric 
elementary excitations with complex eigenfrequencies.
Moreover, their long-term nonlinear behavior after random perturbations shows that
they are semistable: They remain stable under energy-increasing perturbations
but decay under energy-decreasing perturbations.
This is in contrast with the results obtained above for the topologically protected
$(1,1,0)$ embedded solitons. In this case, the existence of an integer-quantized 
topological charge makes the solitons fully stable and thus particularly amenable 
to experimental observation.

We propose the experimental realization of these $(1,1,0)$ embedded solitons 
through a three-stage procedure along the lines of Ref. \cite{Eiermann2004}, by 
using a crossed dipole trap configuration. As a first stage, a (dynamically stable) 
single charged vortex has to be nucleated in the atomic cloud, and to this end a 
2-photon stimulated Raman process with a Laguerre-Gaussian beam along the 
lattice (to be ramped later) could provide the corresponding quantum $\hbar$ of 
orbital angular momentum per atom \cite{Andersen2006}. Next, following 
Ref. \cite{Eiermann2004}, the lattice can be adiabatically ramped up and the atomic 
cloud released on the associated waveguide. Finally, in order for the system to reach 
the boundary of the Brillouin zone from which the gap solitons originate, the lattice has to be 
boosted up to a quasimomentum value of $\pi/d$. By choosing the proper number 
of atoms for the system parameters, the resulting vortex state could be found either 
in a band gap (as a gap soliton) or in an energy band with zero 
angular momentum, thus realizing an embedded soliton.

%\section*{Conclusion}
%\vspace*{0.8cm}
\vspace*{22pt}
{\noindent \Large \textbf{Conclusion}}

\noindent
In conclusion, we have (numerically) obtained for the first time stable 3D embedded 
solitons.
In particular, we have shown that single-component BECs in 1D optical lattices support 
continuous families of stable symmetry-protected 3D embedded solitons in a region of the 
parameter space that is readily accessible with current experimental capabilities.
By imprinting a vortex in the condensate, one can generate robust 
topologically-protected embedded solitons, which may open up a way for the first 
experimental realization of 3D (matter-wave) embedded solitons as well as for
monitoring the gap-soliton to embedded-soliton transition.

%\bibliography{sample}

%\subsection*{\large Acknowledgements}
%\vspace*{0.8cm}
\vspace*{25pt}
{\noindent \large \textbf{Acknowledgements}}

\noindent
V. D. acknowledges financial support from Ministerio de Econom{\'i}a y Competitividad
(Spain) and the Fondo Europeo de Desarrollo Regional (FEDER, EU) under
Grants No. FIS2013-41532-P and FIS2016-79596-P.

%\subsection*{\large Author Contributions}
%\vspace*{0.8cm}
\vspace*{25pt}
{\noindent \large \textbf{Author Contributions}}

\noindent
V.D. proposed the work and wrote the main manuscript text. A.M.M. performed the numerical simulations.
All authors analyzed and interpreted the results and reviewed the final version of the manuscript.
%Both authors contributed equally to this work.
%Must include all authors, identified by initials, for example:
%A.A. conceived the experiment(s),  A.A. and B.A. conducted the experiment(s), C.A. and D.A. analysed the results.  
%All authors reviewed the manuscript. 

%\subsection*{\large Additional Information}
%\vspace*{0.8cm}
\vspace*{25pt}
{\noindent \large \textbf{Additional Information}}

\noindent
Competing Interests: The authors declare no competing interests.

\vspace*{25pt}
{\noindent \large \textbf{Open Access}}

\noindent
This is a preprint of an article published in Scientific Reports. The final authenticated version is available Open Access at: https://doi.org/10.1038/s41598-018-29219-7

\end{document}